# Exploring the Drivers of Information Security Policy Compliance Among Contingent Employees: A Social, Deterrent, and Involvement-Based Approach

Vasty A. Adomako[1], Kaisu Mumuni[1], Eugene M. Akoto[1], and Felix N. Koranteng[1,2*]

[1] *Accra Institute of Technology, Accra, Ghana*
[2] *Eindhoven University of Technology, Eindhoven, The Netherlands*

**Abstract**
As institutions increasingly depend on Information Systems (ISs), ensuring compliance with Information Systems Security Policies (ISSPs) is critical, especially among contingent employees, whose engagement differs from that of permanent staff. This study examines how Subjective Norm, Deterrence (certainty of detection and severity of punishment), and involvement mechanisms (knowledge sharing and collaboration) influence contingent employees' Attitudes Toward ISSPs and, ultimately, their Compliance Intentions. Drawing on data from Ghanaian universities and analyzed using PLS-SEM, the findings confirm that all proposed factors significantly shape attitudes, with knowledge sharing having the strongest effect. Attitude toward ISSPs also strongly predicts compliance intentions. The results support integrating social, cognitive, and collaborative factors into existing ISSP compliance models. Practical implications emphasize fostering inclusive and supportive environments alongside enforcement. This study advances theory and provides a foundation for future research into ISSP behavior among temporary academic staff.

## 1. Introduction

The increasing reliance on Information Systems (ISs) by institutions seeking operational efficiency and competitive advantage is evident across various sectors. Today, many organizations, particularly educational institutions, depend heavily on ISs to perform core functions such as admissions, course registration, and graduation processes [1]. As ISs become integral to institutional operations, securing them is critical to ensuring organizational stability and continuity. In effect, ISs have become the backbone of institutional activities, and their compromise can severely disrupt or even paralyze these organizations [2]. It is therefore essential to proactively manage and mitigate risks that threaten the integrity of these systems.

   Advancements in technology have introduced a range of tools (including firewalls, intrusion detection systems, and antivirus software) to safeguard ISs [3]. While such technologies are effective in countering specific technical threats, an overreliance on them can be misguided. Recent trends indicate that attackers increasingly exploit human vulnerabilities rather than technological ones [4]. In other words, inappropriate user behaviors often serve as entry points for breaches, making users the primary target of many cyberattacks.
In response, institutions have developed Information Systems Security Policies (ISSPs) to regulate user behavior and minimize human-related risks. These policies outline acceptable practices, assign roles and responsibilities, and stipulate consequences for violations [4]. Ideally, ISSPs should empower users to act in ways that reduce system vulnerabilities. However, research consistently reveals a widespread lack of compliance with ISSPs among users [5], [6]. Such noncompliance continues to expose ISs to significant threats.

To address this challenge, several studies have explored the motivational factors influencing users' adherence to ISSPs [1], [5], [6]. While these studies provide valuable insights, most have been conducted in developed countries and primarily focus on permanent employees. There is a noticeable gap in the literature concerning African contexts and the experiences of contingent (i.e., non-permanent or contract-based) employees. Yet, contingent workers often have similar levels of access to institutional ISs and are equally susceptible to targeting by malicious actors. Furthermore, generalizing findings from permanent employees to contingent employees may be problematic, as the psychological contracts and institutional relationships differ significantly between the two groups. As a result, motivational drivers and compliance behaviors may also vary.

This paper seeks to address the identified gap by synthesizing existing literature on ISSP compliance, with a particular focus on the often-overlooked role of contingent employees in the African context. To achieve this, the study adopts a quantitative research approach to empirically assess the effectiveness of key motivational factors that influence contingent employees' intentions to comply with ISSPs. The next section presents a review of relevant literature, followed by a discussion of the theoretical foundation and the formulation of research hypotheses. Subsequently, the research methodology is outlined, after which the results of the empirical analysis are presented. The paper then discusses the key findings in relation to the research objectives, and finally, conclusions are drawn based on the overall insights.

## 2. Literature Review

As institutions increasingly rely on Information Systems (ISs) to support core operations, ensuring the security of these systems has become a critical concern. Existing research highlights both technological and behavioral strategies as essential components in safeguarding ISs [7]. Accordingly, scholars have explored a range of protective mechanisms aimed at mitigating cyber threats.

From a technological standpoint, several countermeasures have been proposed to protect ISs. These include source-end defense mechanisms [8], core-end defense techniques [9], casualty-end protection strategies [10], and probabilistic filter planning [11]. In addition, conventional security tools such as firewalls and cryptographic techniques have proven effective in preserving the confidentiality, integrity, and authenticity of data [12]. However, technological solutions alone are insufficient to ensure comprehensive IS security. Increasingly, scholars recognize the significance of user behavior and human factors in IS protection efforts [7]. Numerous studies have examined the role of individual compliance with ISSPs, identifying a variety of motivational and deterrence mechanisms that influence such behavior. For example, deterrents such as punishment, fear appeals, and perceived threats have been found to significantly affect ISSP compliance [13], [14], [15]. Other studies point to psychological and social factors, such as awareness, leadership, prior experience, self-efficacy, perceived vulnerability, Subjective Norm, and coping mechanisms, as key predictors of compliance behavior [16], [17], [18], [19].

Notably, most of these studies focus on samples drawn from permanent employees in developed countries. For instance, Safa et al. [7] examined ISSP attitudes using participants from four European companies, identifying components of involvement theory (e.g., attachment, experience, and intervention) as influential. However, none of the participants in that study were contingent employees. Similarly, Hwang et. al. [19] found a strong link between information security awareness and compliance behavior among permanent employees in three South Korean organizations.

This paper argues that findings from such studies may not be generalizable to developing countries or to contingent workers. Environmental and cultural differences, such as those highlighted by Hofstede [20], play a critical role in shaping compliance behavior. For example, many African countries score high on power distance, meaning individuals may be more inclined to follow institutional rules and policies compared to their counterparts in low power distance societies, such as much of Europe.

As indicated, these existing studies draw samples from permanent employees as respondents and in developed countries. For example, Safa et al. [7] sampled respondents from four companies in Europe to examine user attitudes toward ISSPs and identified facets of the involvement theory (i.e., attachment, experience, intervention, etc.) as determinants of ISSPs compliance. It is noteworthy that none of the companies or participants from the study were contingent employees. Similarly, Hwang et. al. [19] investigated and found a significant relationship between information security awareness and compliance behavior using permanent employees from three Korean companies. This paper suggests that the differences in environmental factors, such as culture, between developed and developing countries do not permit the generalization of existing findings. According to Hofstede [20], many African countries have a wide power distance; hence, people in these areas are likely to comply with institutional rules and policies, unlike people in European countries.

Additionally, the nature of employment contracts can influence attitudes toward institutional responsibilities. Permanent employees typically have stronger psychological contracts with their organizations, leading to a greater sense of obligation to safeguard institutional assets such as ISs [21]. In contrast, contingent employees may lack the same level of commitment. Research suggests that contingent workers are often more opportunistic, emotionally detached, and less invested in organizational outcomes [22]. As a result, the motivational factors that influence permanent employees' ISSP compliance may be less effective for contingent staff.

Despite their limited inclusion in empirical studies, contingent employees often have similar levels of access to ISs and are equally susceptible to targeting by cybercriminals. Evidence indicates that nearly 50% of data breaches in the United States involve contract or temporary workers [23]. Sharma and Warkentin [4] highlight several cases, such as the Colorado Community Health Alliance breach, which involved temporary personnel. In response to these risks, Guhr [18] advocates for increased research attention on encouraging ISSP compliance among contingent employees. In light of these concerns, this paper aims to empirically examine the unique factors influencing ISSP compliance among contingent employees, particularly within the African context.

## 3. Theoretical Foundations and Conceptual Model

This paper draws from several behavioral theories to explain ISSPs compliance behavior of contingent employees. The paper adopts the Theory of Planned Behavior (TPB) [24] as the fundamental theory. TPB is a very prominent theory in IS research and has been used by extant studies to explain technology use behavior. TPB posits that people form intentions based on their behavior control, attitude (or judgment of the behavior), as well as influence from significant others (respected acquaintances), and these intentions predict their behavior. The TPB, therefore, suggests that people will follow ISSPs when they have the ability to, when they perceive ISSPs positively, and when respected colleagues also follow ISSPs.

Some studies have shown that the TPB is effective in explaining ISSP compliance behavior. Adopting the TPB as a theoretical lens, Ifinedo [25] identified attitude and subjective norm as predictors of employees' intentions to comply with ISSPs. Likewise, Grimes and Marquardson [16] confirmed that Subjective Norm impacts users' intention toward ISSPs. Extant studies have also confirmed the efficacy of TPB in explaining compliance behavior [26], [27], [28]. It is therefore appropriate that this paper adopts the TPB as the fundamental theoretical lens. Despite the TPB's strengths, researchers have noted challenges in accurately measuring perceived behavioral control, with some studies (e.g., Nasir et al., [28]) opting to exclude it from their models. Following this precedent, the current study omits perceived behavioral control and focuses on the relationship between Subjective Norm, attitude, and compliance intention. Specifically, it proposes that Subjective Norm influences contingent employees' attitudes toward ISSPs, and that attitude, in turn, predicts compliance intention.

Further, other concepts may impact contingent employees' Attitude Towards ISSPs. The deterrence theory [29] suggests that individuals are rational and will make decisions to avoid punishment. Specifically, the deterrence theory posits that certainty of detection and severity of punishment influence people's attitudes. Certainty of punishment is the degree of surety of detection of misbehavior, and severity of punishment is the degree of harshness of the penalty for misbehavior. That is, contingent employees will follow ISSPs when they know that non-compliance will be noticed and penalties will be imposed appropriately.

Studies have found deterrence to be effective in influencing ISSP compliance behavior. Safa et. al. [13] identified that certainty of detection and severity of punishment promote a positive attitude towards ISSP compliance. This is corroborated by findings from Rajab and Eydgahi [30]. Related studies have also identified that part-time (i.e., contingent) employees' attitude towards work and performance increases when they face penalties for non-performance [31]. This implies that contingent employees' attitude towards ISSPs may improve when adequate deterrence mechanisms are in place. Based on this, the paper proposes that Certainty of Detection and Severity of Punishment will influence contingent employees' Attitude Towards Compliance.

Next, some studies have argued that individuals do not comply with ISSPs because they do not know or understand the content of the policy [27]. This is reasonable because a person cannot do what he/she does not know. Rocha et. al. [32] argued that the lack of awareness and involvement in ISSP decisions could cause ISSP non-compliance. Indeed, in many organizational settings, contingent employees are often relegated to the background in terms of their involvement in decision-making. It could be possible that these employees may not be preview to the different sections of ISSPs. This may cause them not to comply.

The involvement theory suggests that the involvement manifested through knowledge sharing and collaboration impacts people's attitudes. In information security compliance, this has been supported by findings from Safa [13]. Therefore, this study also proposes that knowledge Sharing and Collaboration will have a significant influence on contingent employees' Attitude Towards ISSPs. The hypothesized relationships are summarized below:

- **H1**: Subjective norms have a significant positive effect on attitudes toward Information Systems Security Policies (ISSPs).
- **H2**: Certainty of detection has a significant positive effect on attitudes toward ISSPs.
- **H3**: Severity of punishment has a significant positive effect on attitudes toward ISSPs.

- **H4**: Knowledge sharing has a significant positive effect on attitudes toward ISSPs.
- **H5**: Collaboration has a significant positive effect on attitudes toward ISSPs.
- **H6**: Attitudes toward ISSPs have a significant positive effect on ISSP compliance intentions.

These hypotheses are graphically illustrated in Figure 1. To test the proposed relationships, the study examines the significance of the corresponding path coefficients within the hypothesized model. The specific procedures and analytical techniques used to assess these relationships are outlined in detail in the Methods section.

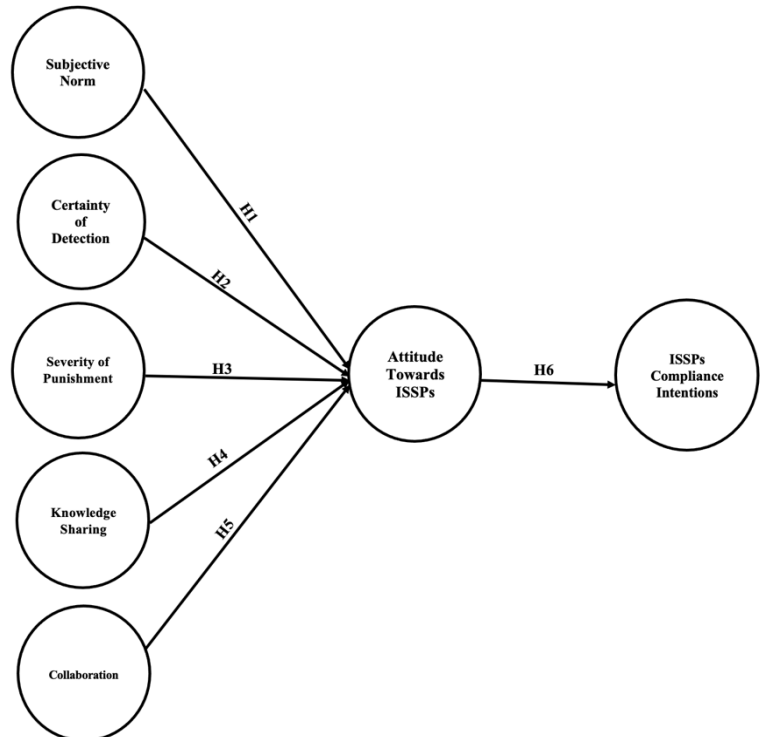

Figure 1. Hypothesized Model

## 4. Method

This section outlines the research methods used to test the hypothesized relationships. It describes the sampling strategy, participant selection process, materials and instruments used, as well as the procedures followed during data collection and analysis

### 4.1. Materials

This study employed a survey-based approach to collect data from participants. An English-based online questionnaire was designed to capture participants' demographics and their perceptions across seven key constructs: (i) Subjective Norm, (ii) Certainty of Detection, (iii) Severity of Punishment, (iv) Knowledge Sharing, (vi) Collaboration, (v) Attitude Towards ISSPs, and (vi) ISSPs Compliance Intentions. To ensure face and content validity, the items used to measure these constructs were adapted from validated instruments in prior research [7], [13], [32], [33].

All questionnaire items were rated using a five-point Likert scale ranging from "Strongly Disagree (1)" to "Strongly Agree (5)".

### 4.2. Procedure

An invitation to participate in the study was distributed via email to individuals affiliated with universities in Ghana. The invitation specifically targeted contingent employees, such as part-time lecturers, teaching assistants, contract staff, and other non-permanent personnel. Interested participants were directed to an online questionnaire via a secure link. Clicking the link indicated their informed consent to voluntarily participate in the study. At the start of the questionnaire, participants were presented with a clear explanation of the study's academic purpose and were assured that their responses would remain confidential and would be used strictly for research purposes. The questionnaire began with a demographic section, which collected information on participants' age, gender, educational background, occupational role, and employment status, specifically to confirm their classification as contingent employees. Following the demographic questions, participants were asked to respond to a series of items measuring key constructs related to ISSP compliance. Responses were captured using a five-point Likert scale ranging from "Strongly Disagree (1)" to "Strongly Agree (5)". Upon completing the questionnaire, participants were prompted to submit their responses and were thanked for their participation. No financial or material incentives were provided for taking part in the study.
.

### 4.3. Participants

A total of 688 contingent employees from various universities across Ghana participated in the study. The sample consisted exclusively of non-permanent university staff, including part-time lecturers, teaching and research assistants, contract-based administrative staff, and other temporary personnel. Eligibility was confirmed through demographic screening at the beginning of the questionnaire. All respondents also reported active affiliation with a university and regular use of their institution's information systems, including learning management systems, and platforms for administrative and communication purposes. Of the 688 participants, approximately 83% (n = 571) identified as male, and 17% (n = 117) identified as female. The mean age of participants was 27 years, with a standard deviation of 5.3 years. The educational background of participants revealed that a majority (48%) held PhD degrees, followed by 34% with master's degrees, 14% with bachelor's degrees, and the remaining 4% holding other qualifications such as diplomas or postgraduate diplomas. In terms of occupational roles, 38% were faculty members or researchers, 33% were teaching or research assistants, while the remaining 29% included contract-based administrative staff, ICT officers, and other temporary university-affiliated professionals.

## 5. Analysis and Results

To evaluate the proposed relationships in the hypothesized model, the study employed Partial Least Squares Structural Equation Modeling (PLS-SEM). This technique was selected due to its robustness against multivariate non-normality and its strength in handling complex, predictive models [34]. Unlike covariance-based approaches, PLS-SEM emphasizes prediction-oriented modeling, making it particularly suitable for exploratory research with a focus on theory development and practical implications [35]. The use of PLS-SEM enabled the examination of causal relationships among constructs and supported the derivation of meaningful design and managerial insights relevant to enhancing ISSP compliance among contingent university employees.

### 5.1. Measurement Model

The first step in the PLS-SEM analysis involved evaluating the measurement model. The measurement model evaluation assesses the reliability and validity of the constructs and indicators used in the study. Specifically, the hypothesized model was evaluated based on item reliability, internal consistency, convergent validity, and discriminant validity. To establish item reliability, a minimum threshold of 0.70 for indicator loadings was adopted. Indicators with loadings below this value were considered for removal. Internal consistency reliability was assessed using Composite Reliability (CR) and Cronbach's Alpha (CA), both of which exceeded the recommended threshold of 0.70, indicating acceptable construct reliability. Convergent validity was examined using Average Variance Extracted (AVE), with a minimum criterion of 0.50. All constructs met this requirement, confirming that a sufficient proportion of variance was captured by the indicators relative to measurement error. For discriminant validity, the Fornell–Larcker criterion was applied. The square root of each construct's AVE was compared against the correlations with other constructs. Discriminant validity was established where the square root of the AVE for a construct exceeded its correlation with any other latent variable [36]. The results, summarized in Table 1, confirm that the measurement model met all required validity and reliability standards, in line with the recommendations of Hair and Sarstedt [34].

Table 1. Construct Reliability and Validity

| Construct | CA | CR | AVE | SN | CD | SP | KS | CO | ATT | CI |
|---|---|---|---|---|---|---|---|---|---|---|
| Subjective Norm (SN) | 0.841 | 0.892 | 0.675 | 0.822 | | | | | | |
| Certainty of Detection (CD) | 0.810 | 0.868 | 0.623 | 0.376 | 0.789 | | | | | |
| Severity of Punishment (SP) | 0.832 | 0.889 | 0.668 | 0.294 | 0.419 | 0.818 | | | | |
| Knowledge Sharing (KS) | 0.850 | 0.902 | 0.700 | 0.435 | 0.521 | 0.483 | 0.837 | | | |
| Collaboration (CO) | 0.823 | 0.882 | 0.651 | 0.422 | 0.394 | 0.457 | 0.605 | 0.807 | | |
| Attitude Towards ISSPs (ATT) | 0.861 | 0.910 | 0.717 | 0.511 | 0.449 | 0.503 | 0.618 | 0.579 | 0.846 | |
| ISSPs Compliance Intentions (CI) | 0.878 | 0.922 | 0.747 | 0.480 | 0.513 | 0.533 | 0.572 | 0.486 | 0.596 | 0.864 |

### 5.2. Structural Model

After confirming the reliability and validity of the measurement model, the structural model was assessed to test the hypothesized relationships among the constructs. The significance of each proposed path was examined using the bootstrap method with 5,000 resamples. A path coefficient was considered statistically significant if the p-value was less than 0.05. The effect size ($f^2$) was evaluated based on Cohen's criteria, where values of $f^2 < 0.02$ indicate no effect, $f^2 \geq 0.02$ a weak effect, $f^2 \geq 0.15$ a moderate effect, and $f^2 \geq 0.35$ a strong effect. The results (summarized in Table 2 and Figure 2) confirm that all hypothesized relationships are statistically significant and in the predicted positive direction, supporting hypotheses H1 through H6.

**Table 2. Significance of Path Coefficients**

| Hypothesis | Path | Original Sample (O) | Sample Mean (M) | p-value | Effect Size ($f^2$) |
|---|---|---|---|---|---|
| **H1** | Subjective Norm → Attitude Towards ISSPs | 0.352 | 0.355 | 0.000 | 0.078 |
| **H2** | Certainty of Detection → Attitude Towards ISSPs | 0.287 | 0.290 | 0.002 | 0.045 |
| **H3** | Severity of Punishment → Attitude Towards ISSPs | 0.313 | 0.316 | 0.001 | 0.052 |
| **H4** | Knowledge Sharing → Attitude Towards ISSPs | 0.411 | 0.412 | 0.000 | 0.123 |
| **H5** | Collaboration → Attitude Towards ISSPs | 0.269 | 0.268 | 0.004 | 0.039 |
| **H6** | Attitude Towards ISSPs → ISSPs Compliance Intentions | 0.586 | 0.590 | 0.000 | 0.376 |

To elaborate, Hypothesis 1 (H1) proposed that Subjective Norm positively influences Attitude Towards ISSPs. The analysis revealed a statistically significant path coefficient ($\beta = 0.352$, $p < 0.001$) and a weak-to-moderate effect size ($f^2 = 0.078$). This suggests that social pressures or expectations from peers, supervisors, and institutional culture substantially shape the attitudes of university contingent employees toward information security policy (ISSP) compliance. Hypothesis 2 (H2) posited that Certainty of Detection positively influences Attitude Towards ISSPs. The relationship was also supported ($\beta = 0.287$, $p = 0.002$) with a weak effect size ($f^2 = 0.045$). This finding indicates that when employees believe there is a high likelihood of being caught for non-compliance, their attitudes toward following ISSPs become more favorable. Similarly, Hypothesis 3 (H3) proposed that Severity of Punishment positively influences Attitude Towards ISSPs. Results showed a significant path ($\beta = 0.313$, $p = 0.001$) and a weak effect size ($f^2 = 0.052$). This implies that awareness of the consequences or penalties associated with non-compliance can slightly but significantly influence employees' attitudes toward adhering to security policies. Hypothesis 4 (H4) explored whether Knowledge Sharing positively influences Attitude Towards ISSPs. The results strongly supported this hypothesis ($\beta = 0.411$, $p < 0.001$) with a moderate effect size ($f^2 = 0.123$). This suggests that a culture of sharing information, experiences, and good practices significantly enhances favorable attitudes toward ISSPs, likely due to increased awareness and perceived relevance. Hypothesis 5 (H5) examined the impact of Collaboration on Attitude Towards ISSPs. The path coefficient was statistically significant ($\beta = 0.269$, $p = 0.004$) with a weak effect size ($f^2 = 0.039$). This indicates that collaborative efforts among peers and departments contribute positively to employees' attitudes, although to a lesser extent than knowledge sharing. Finally, Hypothesis 6 (H6) proposed that Attitude Towards ISSPs positively influences ISSPs Compliance Intentions. This relationship was strong and highly significant ($\beta = 0.586$, $p < 0.001$) with a large effect size ($f^2 = 0.376$). This confirms that the more favorable an employee's attitude is toward information security policies, the more likely they are to intend to comply, underscoring the pivotal role of attitude in compliance behavior.

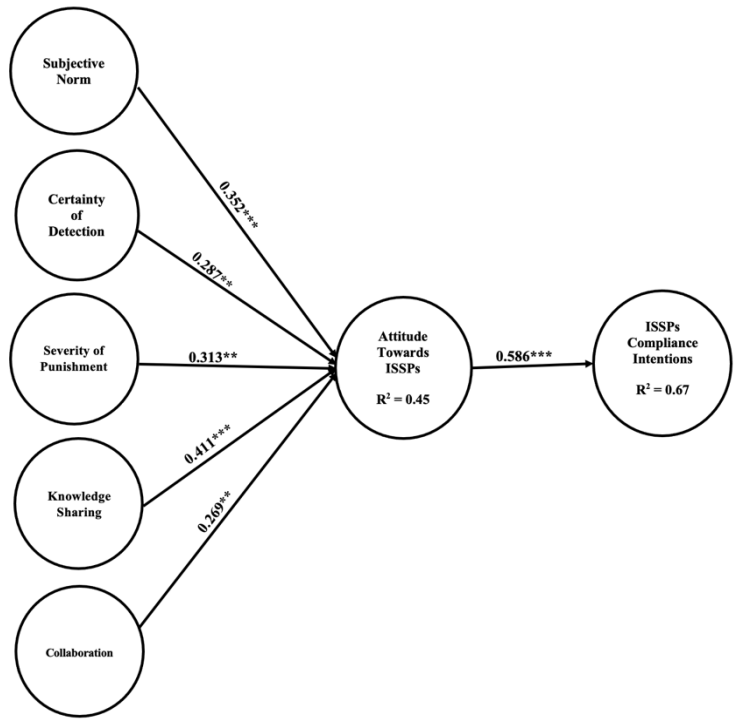

p***<0.00; p**<0.005

Figure 2. Structural Model

## 6. Discussion

This study investigated the effectiveness of salient factors influencing attitudes toward ISSPs and how these attitudes affect compliance intentions among contingent university employees in Ghana. The findings provide relevant insights into how Subjective Norm, certainty of detection, severity of punishment, knowledge sharing, and collaboration shape employees' attitudes toward ISSPs, which in turn significantly influence their intention to comply.

### 6.1. Theoretical Implications

The findings of this study contribute significantly to theoretical understanding in the field of information systems security behavior, particularly regarding the behavioral intentions of contingent employees in academic institutions.

First, the strong positive influence of Subjective Norm on attitudes toward ISSPs reinforces the explanatory power of the Theory of Planned Behavior (TPB) [24], which posits that perceived social pressure plays a central role in attitude formation and behavioral intention. In the context of ISSPs, this finding is consistent with studies by Bulgurcu et. al. [2] and Siponen and Vance[37], who argued that when employees perceive that important referents (e.g., supervisors, coworkers) expect them to comply, their attitudes toward compliance become more favorable. The results validate the argument that even among contingent or non-permanent employees, organizational norms can exert a considerable influence on attitudes, particularly in collectivist or hierarchical environments where conformity to group expectations is emphasized [38]. These findings support the integration of social normative constructs in security compliance models to account for the role of cultural and peer-driven influences in shaping secure behavior.

Second, the observed positive associations between certainty of detection and severity of punishment with attitudes toward ISSPs lend empirical support to Deterrence Theory. While these relationships were marked by weaker effect sizes, their statistical significance confirms that perceptions of potential surveillance and sanctions contribute to forming attitudes that favor compliance. This aligns with findings from Ifinedo [39] and Posey et. al. [40], who argue that even mild awareness of consequences, such as detection mechanisms or disciplinary actions, can reinforce compliance behavior. Notably, this study expands deterrence theory application by demonstrating that its principles remain applicable among temporary workers, who might be assumed to respond less to long-term punitive structures due to their transient roles. This suggests that fear-based deterrents can still be psychologically impactful, especially when security violations are seen as visible, traceable, and consequential within the organizational context.

Third, this study's identification of knowledge sharing and collaboration as strong antecedents of positive attitudes toward ISSPs extends the dominant theories in the information security literature. While TPB and Deterrence Theory center on rational and normative reasoning, our findings suggest that Social Exchange Theory [41] and Organizational Support Theory [42] provide complementary lenses. The positive effect of knowledge sharing confirms prior work by Ahmad and Karim [43] and Hwang et al. [44], who argue that when employees engage in security-related information exchange, their understanding, perceived efficacy, and sense of collective responsibility toward security policies increase. Likewise, the significant role of collaboration is consistent with the assertion by Lebek [45] that interdepartmental cooperation and peer engagement foster trust, shared accountability, and positive security culture, all of which promote favorable policy attitudes. These findings call for an expanded theoretical model that includes relational and community-based constructs, especially within educational institutions where knowledge flow and teamwork are integral to operations.

Finally, the influence of attitude toward ISSPs on compliance intention confirms the central role of attitude as articulated in TPB. The strong path coefficient and large effect size underscore the motivational potency of attitude in the decision-making process. This result is in line with Herath and Rao [14], who found that employees with positive security attitudes are more likely to engage in secure behaviors. It also echoes findings from Siponen and Vance [37], who emphasized the mediating role of attitude in transmitting the effects of both personal beliefs and external influences on behavioral intentions.

### 6.2. Practical Implications

From a practical standpoint, the findings offer clear directions for university administrators, information security managers, and policy designers in improving ISSP compliance, especially among contingent employees such as adjunct faculty, contract researchers, and part-time administrative staff. First, the strong role of Subjective Norm implies that leveraging peer influence and managerial expectations can be an effective compliance strategy. Institutions can foster a security-conscious culture by using role models, departmental champions, and social proof strategies that emphasize collective responsibility for information security.

Second, although the effects of certainty of detection and severity of punishment were relatively weaker, they were still significant. This suggests that institutions should not solely rely on punitive measures, but must maintain visible, consistent, and fair enforcement practices.

Regular communication about monitoring mechanisms and consequences for non-compliance can enhance deterrent perceptions without fostering fear or distrust. Third, the substantial impact of knowledge sharing on attitudes suggests that educational interventions and peer-learning initiatives can be highly effective. Rather than relying solely on formal training, institutions should create opportunities for contingent staff to share security tips, incident experiences, and policy interpretations through workshops, newsletters, and discussion forums. Such knowledge exchange not only improves understanding but also fosters collective engagement. Similarly, collaborative practices should be embedded in institutional procedures. When contingent employees are involved in policy formulation, system testing, or implementation feedback, they are more likely to develop a sense of ownership and responsibility. Encouraging cross-functional collaboration—especially between IT teams and academic departments—can bridge gaps in policy awareness and operational relevance.

Finally, the critical influence of attitude toward ISSPs on compliance intention signals the importance of attitude management in organizational security strategies. Administrators should monitor employee sentiment toward ISSPs and adjust communication, support, and incentives accordingly. Positive framing of policies, highlighting their role in protecting personal and institutional data, and showing appreciation for compliant behavior can strengthen positive attitudes and drive compliance.

## 7. Conclusion and Future Work

This study investigated the determinants of information security policy (ISSP) compliance intentions among contingent employees in Ghanaian universities, focusing on both deterrent and socio-organizational factors. Drawing from the Theory of Planned Behavior (TPB), Deterrence Theory, and Social Exchange Theory, the research found that Subjective Norm, certainty of detection, severity of punishment, knowledge sharing, and collaboration significantly influence employees' attitudes toward ISSPs. Among these, knowledge sharing had the strongest influence, highlighting the critical role of peer-based learning and communication in shaping security behaviors. Attitude toward ISSPs emerged as the most influential predictor of compliance intentions, confirming its central role in behavioral frameworks. These findings provide theoretical insight into how cognitive, normative, and relational factors interact to influence security behavior, particularly within the context of temporary university staff who may have limited institutional affiliation.

Practically, the results suggest that security interventions should go beyond enforcement to include strategies that foster collaboration, open communication, and social reinforcement of compliance behaviors. Future research should extend these findings by incorporating longitudinal data to assess changes over time, examining the moderating effects of employment type or institutional policies, and integrating other contextual variables such as organizational climate and leadership commitment to develop more comprehensive and inclusive ISSP compliance models.